\documentclass[preprint,12pt]{elsarticle}

\usepackage{graphicx}

\usepackage{amssymb}

\journal{New Astronomy}

\begin{document}

\begin{frontmatter}

\title{Spectra of dynamical Dark Energy cosmologies from constant--$w$ models}

\author[mib,infn]{Luciano Casarini}
\ead{luciano.casarini@mib.infn.it}

\address[mib]{Department of Physics G.~Occhialini, Milano Bicocca
University, Piazza della Scienza 3, 20126 Milano, Italia}
\address[infn]{I.N.F.N., Sezione di Milano}

\begin{abstract}
WMAP5 and related data have greatly restricted the range of acceptable
cosmologies, by providing precise likelihood ellypses on the the
$w_0$--$w_a$ plane. We discuss first how such ellypses can be
numerically rebuilt, and present then a map of constant--$w$ models
whose spectra, at various redshift, are expected to coincide with
acceptable models within $\sim 1\, \%$.
\end{abstract}


\end{frontmatter}


\section{Introduction}
\label{intro}

One of the main puzzles of cosmology is why a model as $\Lambda$CDM,
implying so many conceptual problems, apparently fits all linear data
in such unrivalled fashion \citep{astier,riess,komatsu}.

It is then important that the fine tuning paradox of $\Lambda$CDM is
eased in cosmologies where Dark Energy (DE) is a self--interacting
scalar field $\phi$ (dDE cosmologies), with no likelihood downgrade
\citep{colombo,lavacca}. Unfortunately, however, only cosmological observables
can provide information on the form of the self--interaction
potential, even though several researchers incline to priviledge
potentials allowing tracking solutions.

When aiming to obtain information on the physical potential, the basic
observable is however the evolution of the DE scale parameter, $w(a)$.
Here $a = 1/(1+z)$ is the scale factor of a spatially flat metric
\begin{equation}
ds^2 = c^2 dt^2 - a^2(t) d\ell^2~,
\label{metric}
\end{equation}
while $z$ is the redshift.

The analysis of available data, made by the WMAP team \citep{komatsu},
was able to constrain the coefficients $w_0$ and $w_a$ in the
expression
\begin{equation}
w(a) = w_0 + (1-a)\, w_a~,
\label{wa}
\end{equation}
putting again into evidence that a model with $w \equiv -1$ is close
to top likelihood, but also stressing a preference for the phantom
area ($w < -1$), which is hardly consistent with current tracking
potentials.

The main tool, to go beyond these constraints on the $w(a)$ law, will
be tomographic shear analyses \citep{refregier}, able to reconstruct
the spectrum of density fluctuations at various $z$'s, with a
precision approaching 1$\, \%$ \citep{huterer}.

This work is therefore focused on the relevance of spectral
predictions, and aims at providing a tool to ease the determination of
model spectra. Within this context, it is essential to outline the spectral
equivalence criterion (SEC). 

It has been noted since a few years \citep{white,linderW} that the
density fluctuation spectra $P(k) = \langle |\delta \rho/\rho |^2
\rangle$, up to $k=3\, h$Mpc$^{-1}$, essentially depend on the
distance from the LSB (Last Scattering Band). This was first verified
at $z=0$ \citep{linder} and then tested at higher $z$ \citep{CaMaB}.
$N$--body simulations show that the SEC is however true when model
parameters are suitably tuned.

Let us be more specific on this point: When a given $w(a)$ is
considered, it is easy to determine the comoving distance of the
LSB. Keeping then the same values of $\Omega_{b,c}$ and $h$, we can
seek a constant--$w$ cosmology with the same distance from the LSB;
\citep{linder} tested that the spectra of these two cosmologies
coincide (within $1\, \%$). They also saw that, when $z \neq 0$ values
are explored, spectral discrepancies are mostly greater, in the range
of a few percents.
\begin{figure}
\vskip -2.4truecm
\begin{center}
\includegraphics[width=10.cm]{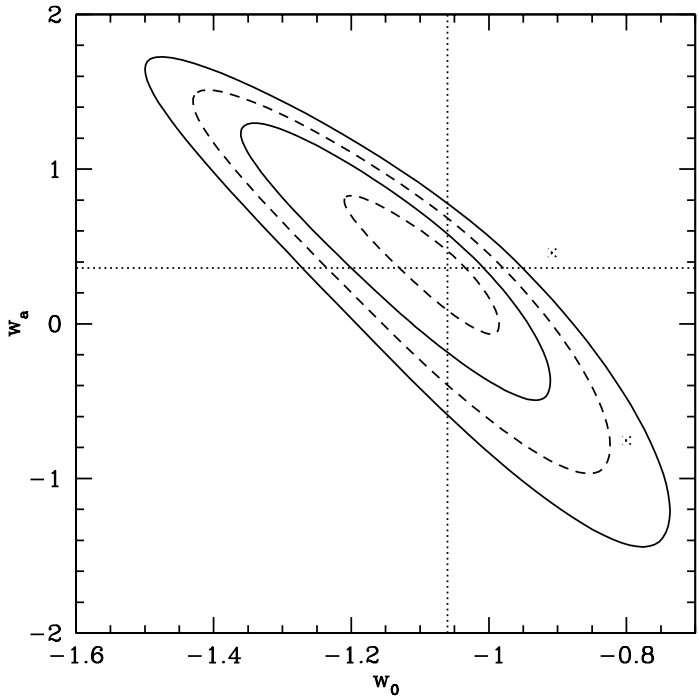}
\end{center}
\vskip -.6truecm
\caption{1-- and 2--$\sigma$ curves, yielding the marginalized model
likelihood on the $w_0$--$w_a$ plane, as obtained from the Lambda NASA
site. The reproduction device is detailed in the text.  Also 0.5-- and
1.5--$\sigma$ curves are provided. Crosses indicates models for which
the SEC was explicitly tested at various $z$'s. }
\vskip -.3truecm
\label{elly}
\end{figure}

The SEC however works also at $z \neq 0$, provided that the distance
between such $z$ and the LSB is evaluated and a constant--$w$
auxiliary model is defined, with an equal distance between $z$ and the
LSB. The assigned cosmology and the auxiliary model must also have
equal $\Omega_{b,c}$, $h$ {\it at $z=0$}, while $\sigma_8(z=0)$ must
be tuned in order that, at $z$, the {\it r.m.s.}  density fluctuations
of the two models, on the $8\, h^{-1}$Mpc scale, coincide \citep{CaMaB}.

Let us outline, in particular, that the SEC does not require that the
auxiliary model shares the values of $\Omega_{b,c}$ and $h$ at the
assigned $z \neq 0$. In fact, by multiplying the critical density
definition by $\Omega_{c,b}$, one has
\begin{equation}
\omega_{c,b} \propto \Omega_{c,b} H^2 = (8\pi G/3) \Omega_{c,b} \rho_{cr}
= (8\pi G/3) \rho_{c,b} \propto a^{-3}
\end{equation}
so that the assigned cosmology and the auxiliary model share the
values of $\omega_{c,b}$ at any $z$, and this is enough.

This comes as no surprise, however: most linear feature, {\it e.g.}
BAO's, essentially depend just on $\omega_{b,c}$.

This recipe, however, prescribes a different constant--$w$ at any $z$
and is somehow curious that, starting from the assigned $w(a)$, one
builds a $w_{eff}(a)$ law, completely different from it. In a similar
way, one can build a $\sigma_{8;(z=0)}$ dependence from $z$, in order
that auxiliary models be fairly normalized at any $z$. Examples of
$w_{eff}(z) $ and $\sigma_{8;(z=0)}(z)$ are provided by \citep{CaMaB}.

The scope of this work, instead, amounts to applying the SEC to models
with $w(a)$ of the form (\ref{wa}) and consistent with data, exploring
decreasing likelihood ellypsoids.

The plan of the paper is as follows: In the next Section we shall
discuss how one can rebuild the likelihood ellypsoids on the
$w_0$--$w_a$ plane.  This Section defines the parameter $t$ used in
the plots which are one of the results of this work. In Section 3 we
shall then discuss such plots. Finally, Section 4 will be devoted to
drawing our conclusions.

\section{How to reproduce likelihood curves with B\'ezier paths}
\label{}

Solid lines in Figure \ref{elly} reproduce the likelihood ellypsoids
on the $w_0$--$w_a$ plane of Figure 14 in \citep{komatsu}. It is
significant to explain how such reproduction is obtained, as the
technique used is also essential to build the forthcoming Figures,
where a parameter $t$ appears, which are one of the results of this
work. In a sense, in this way we pass from an eulerian $w_0$--$w_a$
description to a lagrangian $t$ description, also stressing the
simmetries of the likelihood. 

Dashed lines in Figure \ref{elly} approximately yield 0.5-- and
1.5--$\sigma$'s; they are defined as the locus of points of equal $t$
exactly at the center of the intervals between top likelihoods and 1
$\sigma$ or 1-- and 2--$\sigma$'s.
   
The technique we shall describe to draw curves on a plane, named after
{\it B\'ezier}, is largely used in vector graphics to model smooth
curves which can be scaled indefinitely, without any bound, by the
limits of rasterized images. 

For instance, imaging systems like PostScript, use cubic B\'ezier
curves:
\begin{equation} 
\label{bezier}
{\bf B}(u)=(1-u)^3 {\bf P}_0 + 3(1-u)^2 u {\bf P}_1+3 (1-u)u^2 {\bf
P}_2+u^3 {\bf P}_3~~ u \in [0,1].
\end{equation}
The vector {\bf B} runs on the $w_0$--$w_a$ plane, describing a curve
when $u$ varies from 0 to 1. The curve is fixed by the positions of
the points $P_k$ (k=0,...,3). In eq.~(\ref{bezier}), ${\bf P}_k$
indicates a vector ending on the very ${P}_k$ point. The curve starts
at $ P_0$ and is initially directed toward $P_1$; however, it bends
soon, owing to the setting of all other points; for $u=1$, it meets
$P_3$, its final direction being set by $ P_2$. Clearly, it hits
neither $P_1$ nor $P_2$: these points only provide directional
information, while the distance $|{\bf P_0}$--${\bf P_1}|$ tells us how
persistently the curve moves towards $P_1$; an analogous effect, at the
end of the run, is fixed by the distance $|{\bf P_3}$--${\bf P_2}|$.

Quadratic and cubic B\'ezier curves are mostly used; when more complex
shapes are needed, rather than making recourse to higher degree
curves, numerically expensive to evaluate, lower degree B\'ezier
curves are patched together. In our specific case, the ellypsoid is
shared in four paths, each described by an expression
(\ref{bezier}). Let us then label the paths with $i$; we can follow
the whole ellypsoid with a single parameter $t \in [0,1]$ by labelling
the paths with an index $i$ ($i=1,..,4$) and setting $u=4t-i+1$;
accordingly, it shall be
\begin{equation} 
\label{ut}
i-1 < 4t < i~~~~~~~~~~~~~{\rm in ~the}~i - {\rm th~path.}
\end{equation}
Using the PostScript file of Figure 14 in \citep{komatsu}, we can
easily obtain the $x,y$ coordinates of the $P_k$ points of each
path. In table 1
we report such points for 1--$\sigma$ and 2--$\sigma$ ellypsoids,
converted into $w_0$--$w_a$ units.

\begin{table}[htbp]
\begin{center}
\begin{tabular}{|l||c|c||c|c||c|c||c|c||}
\hline
  & $x_0$ & $y_0$ & $x_1$ & $y_1$ & $x_2$ & $y_2$ & $x_3$ & $y_3$\\
\hline
i & & & & & & & & \cr
\hline
1 & -1.3593 & 1.2597 & -1.3765 & 1.1038 & -1.2247 & 0.4538 & -1.1387 & 0.1068\\
\hline
2 & -1.1387 & 0.1068 & -1.0528 & -0.2402 & -0.9333 & -0.6217 & -0.9145 & -0.4527\\
\hline
3 & -0.9145 & -0.4527 & -0.8942 & -0.2704 & -0.9542 & 0.1407 & -1.0326 & 0.4713\\
\hline
4 & -1.0326 & 0.4713 & -1.1318 & 0.8893 & -1.3384 & 1.4488 & -1.3593 & 1.2597\\
\hline
\end{tabular}
\end{center}
\vskip -1.truecm
\label{tsig1}
\end{table}
\begin{table}[htbp]
\begin{center}
\begin{tabular}{|l||c|c||c|c||c|c||c|c||}
\hline
 & $x_0$ & $y_0$ & $x_1$ & $y_1$ & $x_2$ & $y_2$ & $x_3$ & $y_3$ \cr
\hline
i &  &  &  &  &  &  &  & \\
\hline
1 & -1.4967 &  1.6957 & -1.5229 &  1.5197 & -1.3801 &  0.8544 & -1.1351 & -0.2619 \\ 
\hline
2 & -1.1351 & -0.2619 & -0.9558 & -1.0784 & -0.7862 & -1.6310 & -0.7469 & -1.3819 \\
\hline
3 & -0.7469 & -1.3819 & -0.7039 & -1.1091 & -0.8015 & -0.2921 & -0.9625  & 0.4120 \\
\hline
4 & -0.9625 & 0.4120 & -1.1235 &  1.1160 & -1.4676 &  1.8905 & -1.4967 &  1.6957 \\
\hline
\end{tabular}
\end{center}
\caption{Points defining the 4 cubic B\'ezier expressions yielding the
1--$\sigma$ and 2--$\sigma$ curves (upper and lower table, respectively).}
\label{tsig2}
\end{table}

These coordinates were used to produce Figure \ref{elly}. This
description is however essential also to introduce the parameter $t$,
which shall be used in the forthcoming plots.
\begin{figure}[htbp]
\begin{center}
\includegraphics[width=10cm]{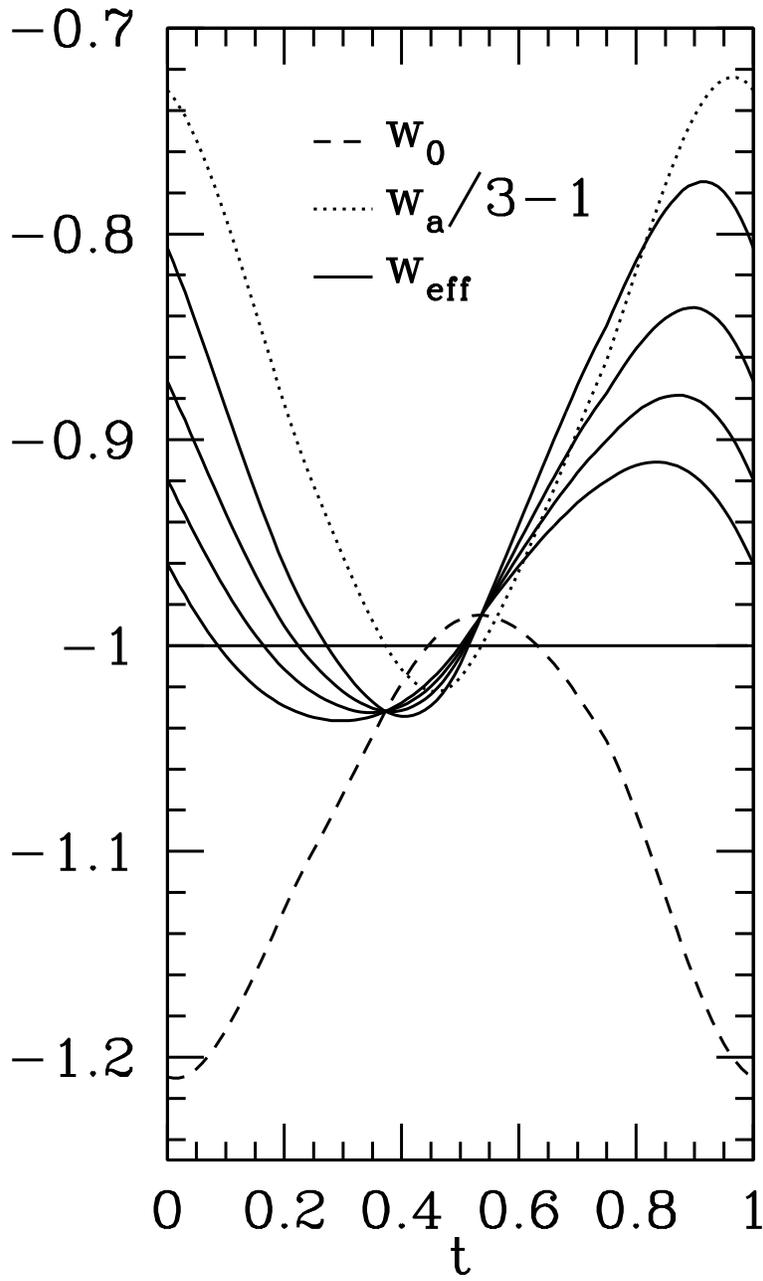}
\end{center}
\caption{This plots gives the state parameters for constant--$w$
models (solid lines), yielding the same distance from the LSB of
$w_0$--$w_a$ models at 0.5 $\sigma$'s from the top likelihood
cosmology; the 4 solid line refer to $z = 0$, 0.5, 1, and 2~.}
\label{f05}
\end{figure}

\section{Results}
\label{}

In fact, by letting the parameter $t$ to run, we move along each curve
of Figure \ref{elly}, and also define the {\it so--called}~
0.5--$\sigma$ and 1.5--$\sigma$ curves.

In Figure \ref{f05} we report the variations of $w_0$ (dashed curve)
and $w_a$ (dotted curve), when $t$ runs from 0 to 1 (we plot $w_a/3-1$
instead of $w_a$, to allow wider ordinate spacing). Each $t$ value,
then, yields a $w_0$--$w_a$ couple, defining a model at 0.5--$\sigma$
from top likelihood. In the same Figure, we also report how the
constant state parameter $w_{eff}$ depends on $t$, at various
redshift, for a model yielding the same distance of 0.5--$\sigma$
models from the LSB. The redshift values considered are $z = 0$, 0.5,
1, and 2~. When $z$ increases, the ordinate interval spanned by
$w_{eff}$ values becomes wider; this is true at any number of
$\sigma$'s and allows to individuate the solid curve referring to each
$z$.

\begin{figure}[htbp]
\begin{center}
\includegraphics[width=10cm]{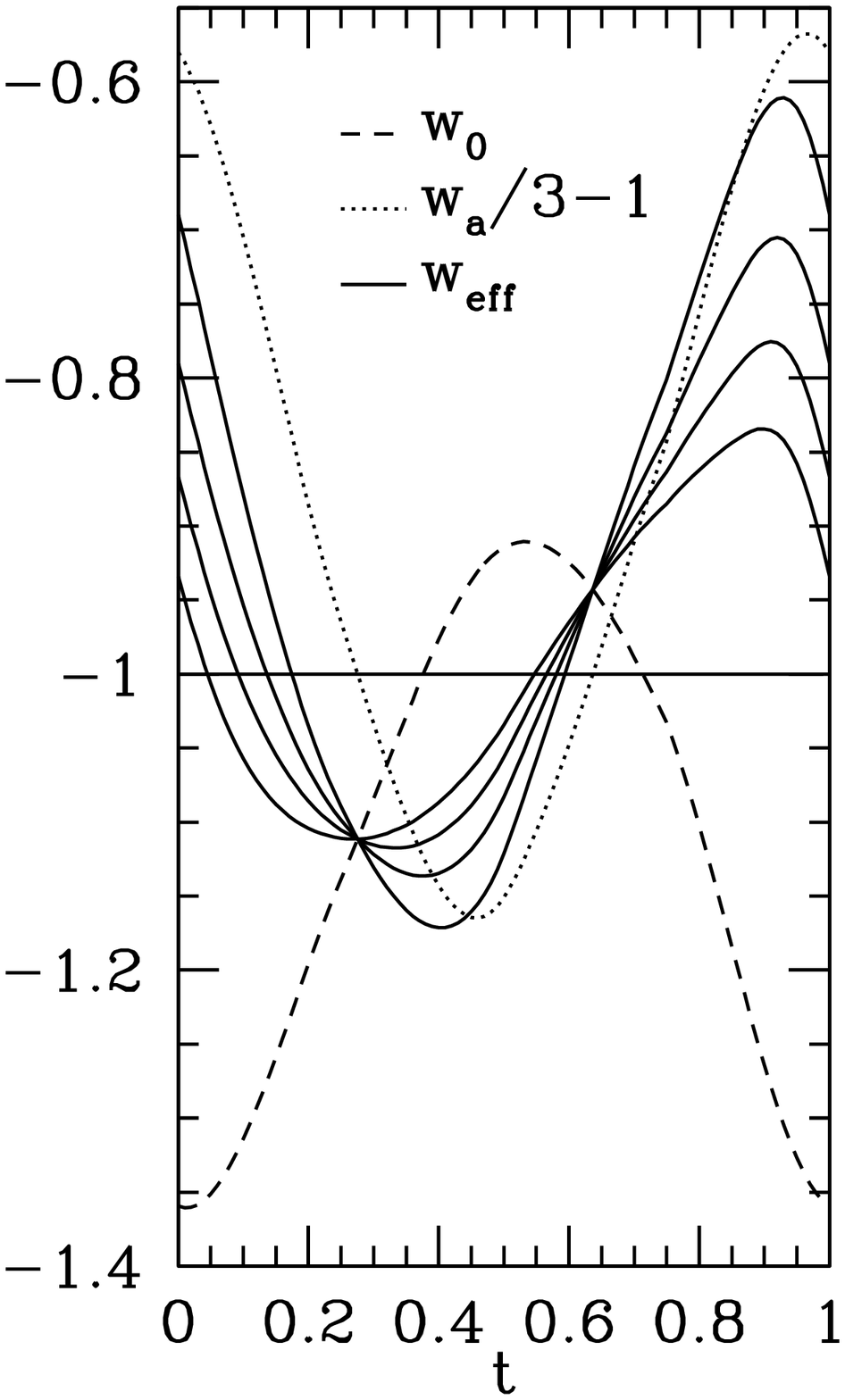}
\end{center}
\caption{As Figure \ref{f05}, for 1--$\sigma$ models.}
\label{f1}
\end{figure}
\begin{figure}[htbp]
\begin{center}
\includegraphics[width=10cm]{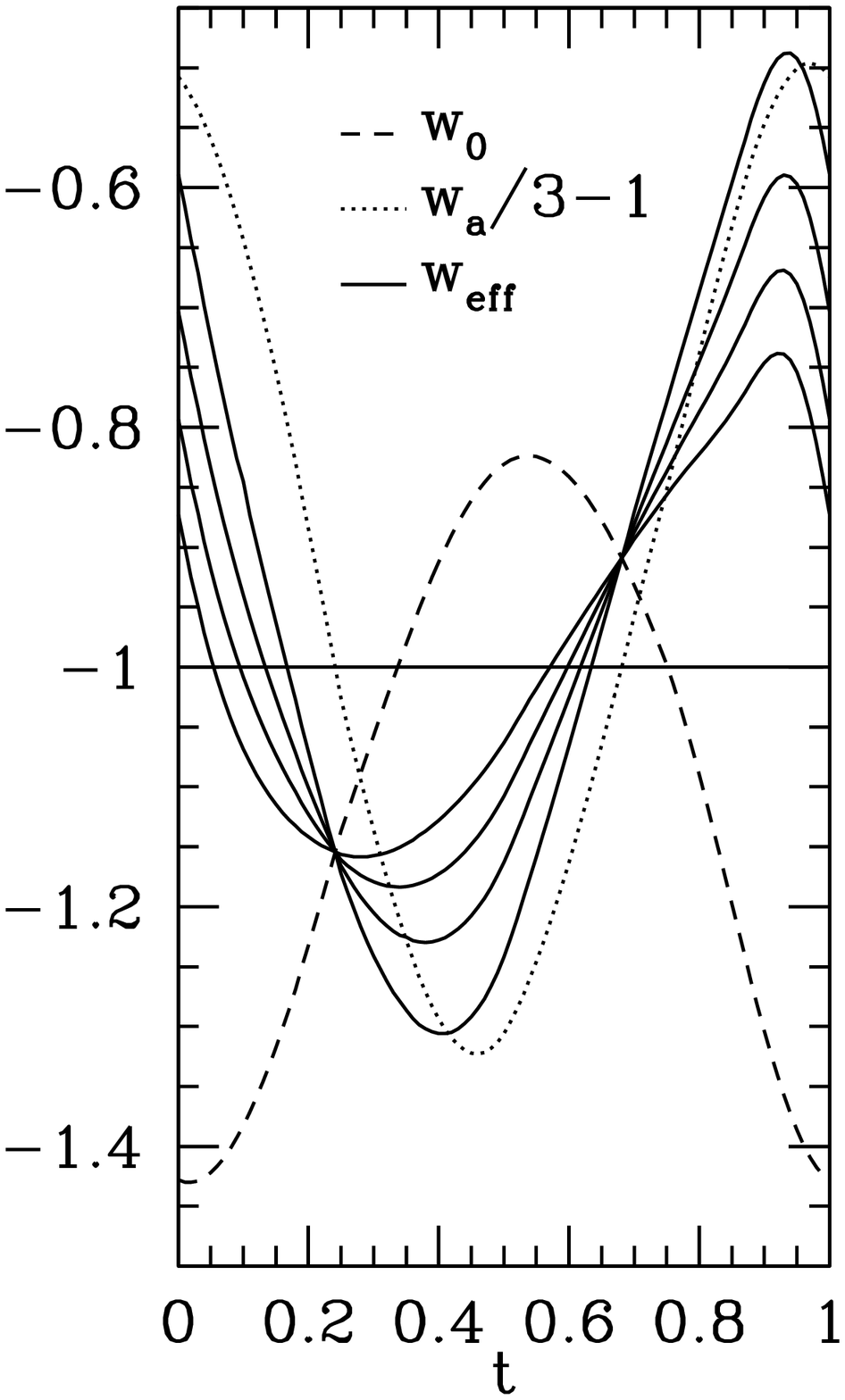}
\end{center}
\caption{As Figure \ref{f05}, for 1.5--$\sigma$ models.}
\label{f15}
\end{figure}
\begin{figure}[htbp]
\begin{center}
\includegraphics[width=10cm]{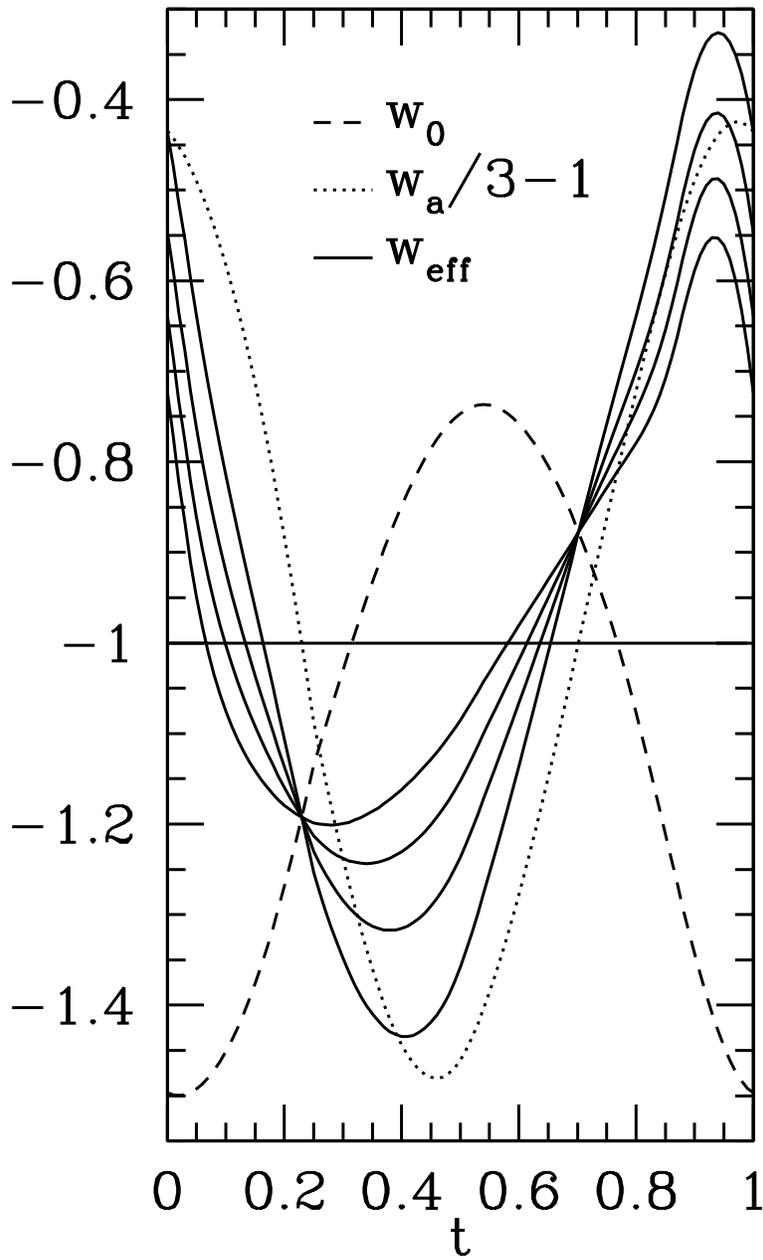}
\end{center}
\caption{As Figure \ref{f05}, for 2--$\sigma$ models.}
\label{f2}
\end{figure}

Figures \ref{f1}--\ref{f2} provide analogous results for models at 1,
1.5 ad 2 $\sigma$ from the top--likelihood cosmology.

These figure are obtained by taking $\omega_c=0.228$,
$\omega_b=0.046$, $h=0.71$.

Let us notice that solid curves converge and meet dashed ones for $t$
values where $w_a$ vanishes. These points also set a transition
between the $t$--intervals where $w_{eff}$ increases or decreases with
$z$.

The top likelihood, as is known, corresponds to a {\it phantom}--DE
model. At $0.5~\sigma$'s from it most models have $w_0 < -1$ but, also
a large fraction of the few models with $w_0 > -1$ correspond to
$w_{eff} < -1$. On the contrary, there are quite a few models,
characterized by $w_0 < -1$ whose spectra are equivalent to
constant--$w$ models with $w_{eff} > -1$.

The $t$ interval where $w_0 < -1$ becomes wider when a greater number
of $\sigma$'s is considered. On the contrary, the width of the
intervals characterized by $w_0 < -1$ and $w_{eff} > -1$ does not
change much with the number of $\sigma$'s.

\section{Conclusions}
\label{conclusions}

When tomographic cosmic shear data will become available, an
inspection on DE nature will surely start from comparing them with the
predictions of constant--$w$ cosmologies.

As soon as data will become more refined, it will be possible to bin
them, discriminating among the $w$ values best fitting data at various
redshift.

It is quite possible that these inspections yield $w$ values
compatible with a constant, all through the redshift range explored.
As it is possible that such value is compatible with $-1$, so
vanifying the efforts to improve our knowledge on DE nature.

Let us suppose that, instead, data analysis is consistent with the
same $\omega_{b,c}$ values in all bins, but require different $w$'s in
different bins. 
\begin{figure}
\begin{center}
\includegraphics[width=10cm]{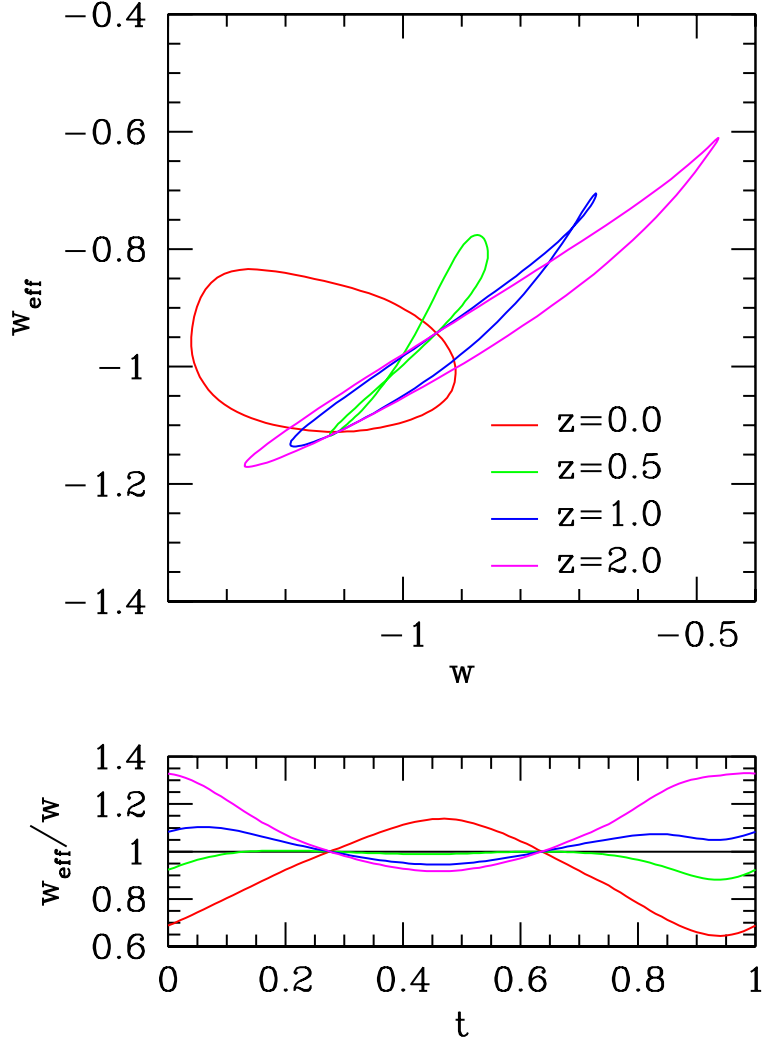}
\end{center}
\caption{Bias in observational $w$ values (named here $w_{eff}$), in
respect to the physical value of the state parameter, at each
redshift. The points in the curves of the upper frame refer to
different models at 1--$\sigma$ from the top--likelihood cosmology,
discriminated~by different $t$ values. In the lower frame the ratio
$w_{eff}/w$ is explicitly shown as a function of $t$~.}
\label{tw21}
\end{figure}
\begin{figure}
\begin{center}
\includegraphics[width=10cm]{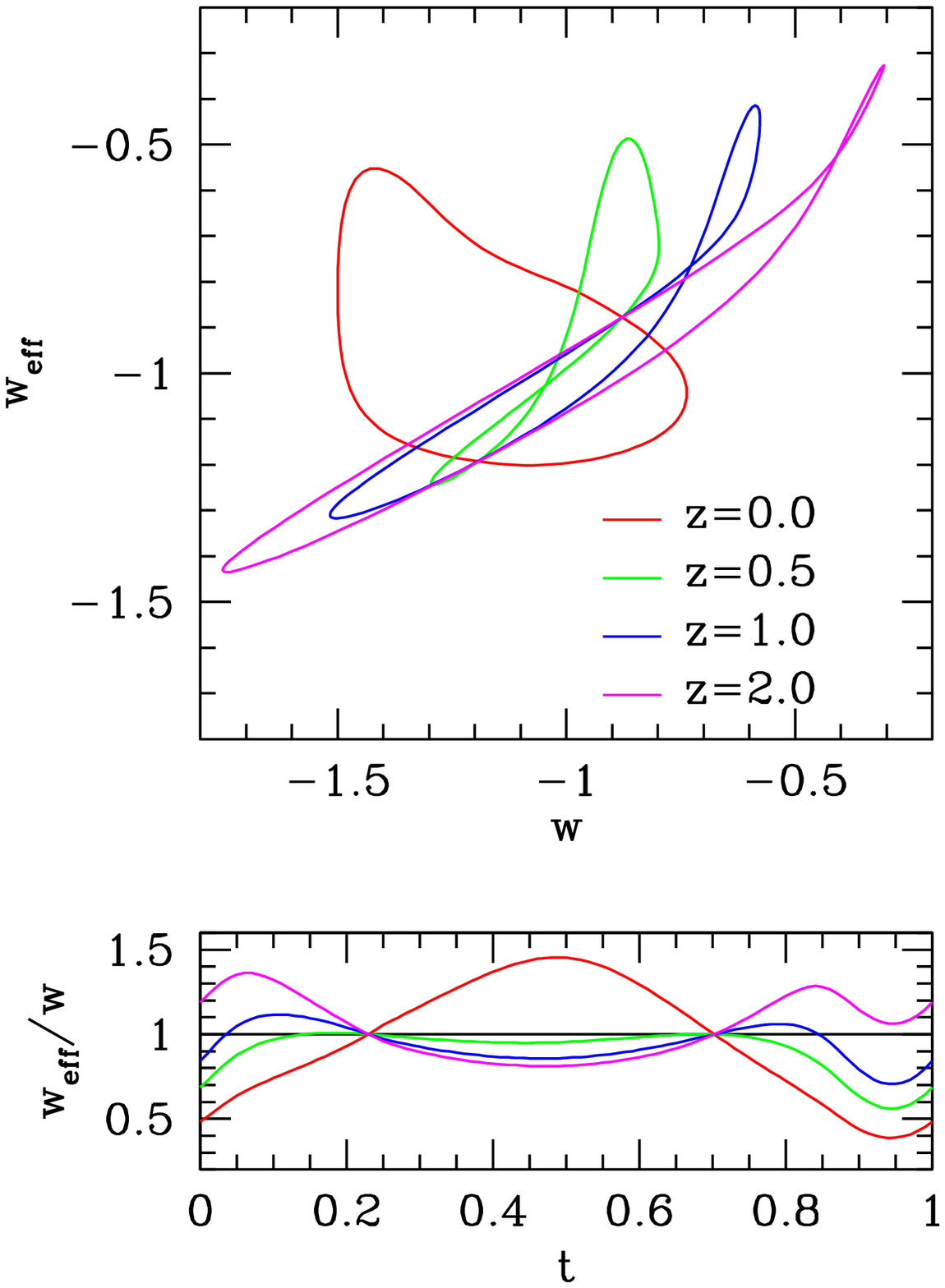}
\end{center}
\caption{As Figure \ref{tw21} for 
models at 2--$\sigma$ from the top--likelihood cosmology.}
\label{tw22}
\end{figure}
We wish to outline here a major danger that data analysis could meet
in this welcome case. As a matter of fact, one could be tempted to
conclude that the $w(z)$ dependence found is the physical scale
dependence of the DE state parameter. Unfortunately, this could be
badly untrue.

In fact, cosmic shear spectra can be easily translated into
fluctuation spectra, so that observational values of $w(z)$ would
correspond to $w_{eff}(z)$, not to the physical $w(z).$

How different the two behaviors can be is already represented by
Figures 2--5 in this paper. But Figures \ref{tw21} and \ref{tw22}
further illustrate this point. They show what is $w_{eff}$ when $w$
has a given value, for models laying on the 1--$\sigma$ or 2--$\sigma$
curves.  Different colors in each plot correspond to different
$z$'s. In the lower frame, the $w_{eff}/w$ ratio is also shown, as a
function of $t$.  Even in the most furtunate cases, provided that the
{\it true} cosmology is not a constant--$w$ one, discrepancies are
hardly below 10$\, \%$ and are greatest at $z$ approaching zero.

Another point concerning model fitting is that a systematic mapping of
(reasonable) dDE cosmologies, in order to fit future data, is
apparently unnecessary. At each $z$, in fact, there will be a
constant--$w$ model able to fit any dDE cosmology. dDE will then be
revealed by the $w$ dependence on $z$, as above outlined. At present,
well approximated analytical expressions of $P(k)$, at different $z$,
are provided by the so--called {\it Halofit} formulae, holding for
$\Lambda$CDM cosmologies \citep{smith}. Some attempt to generalize {\it
Halofit} to constant $w \neq -1$ were also performed \citep{mcdonald},
but they do not cover the desired parameter ranges. Our conclusion is
that it will be important to extend {\it Halofit} to constant--$w$
cosmologies, for the whole range of (reasonable) cosmological
parameters; this will enable us to fit future cosmic shear data
without any substantial restriction on the $w(a)$ behavior.

\section{Acknowledgments}
\label{thanks}

Thanks are due to Silvio Bonometto for a number of discussions, for
providing some numerical tools and discussing the final text of this
article.








\section*{References}
\newcommand{\Nature}{{\it Nature\/} }
\newcommand{\ApJ}{{\it Astrophys. J.\/} }
\newcommand{\ApJS}{{\it Astrophys. J. Suppl.\/} }
\newcommand{\MNRAS}{{\it Mon. Not. R. Astron. Soc.\/} }
\newcommand{\PhRv}{{\it Phys. Rev.\/} }
\newcommand{\PhL}{{\it Phys. Lett.\/} }
\newcommand{\JCAP}{{\it J. Cosmol. Astropart. Phys.\/} }
\newcommand{\AeA}{{\it Astronom. Astrophys.\/} }
\newcommand{\etall}{{\it et al.\/} }
\newcommand{\arXiv}{{\it Preprint\/} }

\end{document}